\documentclass[fleqn,usenatbib]{mnras}

\usepackage{newtxtext,newtxmath}

\usepackage[T1]{fontenc}

\DeclareRobustCommand{\VAN}[3]{#2}
\let\VANthebibliography\thebibliography
\def\thebibliography{\DeclareRobustCommand{\VAN}[3]{##3}\VANthebibliography}

\usepackage{graphicx}	
\usepackage{amsmath}	
\usepackage[flushleft]{threeparttable}

\newcommand{\Msun}{$M_{\odot}$}

\newcommand{\tess}{\it TESS}

\title[X-ray Activity of M dwarf UFRs]{The puzzling story of flare inactive ultra fast rotating M dwarfs – III. Investigating X-ray Activity}

\author[L. Doyle \& G. W. King et al.]{
Lauren Doyle,$^{1,2}$\thanks{E-mail: lauren.doyle@warwick.ac.uk}\footnote[3]{}
George W. King,$^{3}$\thanks{E-mail: kinggw@umich.edu}\thanks{These two authors contributed equally to this work and should be considered joint first authors.}
Gavin Ramsay,$^{4}$
L\'ia R. Corrales,$^{3}$
Stefano Bagnulo,$^{4}$
\newauthor
J. Gerry Doyle$^{4}$
and Pasi Hakala,$^{5}$
\\
$^{1}$Centre for Exoplanets and Habitability, University of Warwick, Coventry, CV4 7AL, UK \\
$^{2}$Department of Physics, University of Warwick, Coventry, CV4 7AL, UK \\
$^{3}$Department of Astronomy, University of Michigan, Ann Arbor, MI 48109, USA\\
$^{4}$Armagh Observatory and Planetarium, College Hill, Armagh, BT61 9DG, UK \\
$^{5}$Finnish Centre for Astronomy with ESO (FINCA), Quantum, University of Turku, FI-20014, Finland
}

\date{Accepted 2026 April 2. Received 2026 April 1; in original form 2026 February 3}

\pubyear{\the\year}

\begin{document}
\label{firstpage}
\pagerange{\pageref{firstpage}--\pageref{lastpage}}
\maketitle

\begin{abstract}
According to activity-rotation relations, rapid rotators are expected to show high levels of magnetic activity. However, recent studies with {\tess} have found Ultra Fast Rotating (UFR) M dwarfs with periods $<1$\,d displaying low levels of flaring activity. There have been efforts to explore their magnetic field strengths through spectropolarimetric measurements and to assess the potential for binarity. However, neither could fully explain the lack of observed flaring activity despite their rapid rotation. Another avenue for investigation is to measure their coronal emission for signs of supersaturation: an underluminosity in X-rays observed for some rapidly rotating FGK stars. Therefore, in this study, we utilise X-ray observations from {\sl Swift} and {\sl XMM-Newton} of ten M dwarf UFRs with P$_{\rm{rot}}$<1\,d to determine their X-ray luminosities. Overall, we do not find evidence for supersaturation amongst our UFR M dwarf stars, instead determining them to be at the saturated level, or perhaps even enhanced. Therefore, supersaturation seems not to be the main driver behind the reduced level of flaring activity observed in these stars, and the mystery behind the magnetic activity of UFR low-mass stars remains. Additionally, we provide an updated analysis on the long term variability within our sample using {\tess} light curves taken during Cycles 5 and 7. We identify 352 optical flares from our sample with energies between $1.2\times10^{31}$ and $8.7\times10^{34}$~erg. We determine flare rates for each {\tess} cycle, compare them, identifying variations across a 7 year timespan and attribute this to potential activity cycles. 

\end{abstract}




\section{Introduction}
Magnetic fields in low-mass stars are generated through a dynamo process operating in their interiors, giving rise to the observed magnetic activity such as stellar flares, star spots, and coronal/chromospheric emission. The stellar rotation rate is important for setting the level of magnetic activity. Activity has been observed to saturate (e.g. the ratio of the X-ray and bolometric luminosities, $L_{\rm X}/L_{\rm bol} \sim 10^{-3}$) in rapid rotators, and decreases systematically as magnetic braking slows the rotation rate \citep{Skumanich1972, hartmann1987rotation, maggio1987einstein, kiraga2007age, yang2017flaring}. The exact dynamo mechanisms operating in low-mass stars remain incompletely understood, particularly in fully convective stars \citep[$M_{\star} <$0.35~M$_{\odot}$:][]{chabrier1997structure} where the absence of a tachocline suggests their magnetic fields are generated by a distributed $\alpha^{2}$ or turbulent dynamo rather than the solar-like $\alpha \Omega$ mechanism \citep[see][]{chabrier2006large, donati2006large, browning2008simulations, mullan2020transition, yadav2016magnetic}. Intriguingly, X-ray-rotation relationships appear to show similar behaviour for fully convective stars compared to their partially convective counterparts \citep{Wright2016, Wright2018}.

It is generally accepted that the interactions between a star's convective motions and stellar rotation play a crucial role in the generation and maintenance of a magnetic field and the observed magnetic activity. The Rossby number ($R_{o}$), defined as the ratio of stellar rotation period to convective turnover time, provides a physically motivated parameter for characterising stellar dynamos and magnetic activity, with lower Rossby numbers corresponding to stronger activity and the onset of saturation \citep{Noyes1984, Pizzolato2003, Wright2011}. At the lowest Rossby numbers, some rapidly rotating stars exhibit a decline in magnetic activity despite increasing rotation rates, a phenomenon known as supersaturation. This effect has been observed primarily in X-ray emission, where activity levels fall below the saturated plateau for ultra-fast rotators, particularly among young and low-mass stars \citep{Prosser1996, Randich1996, Stkepien2001, jeffries2011investigating, Wright2011}. Several mechanisms have been proposed to explain supersaturation, including centrifugal stripping of the outer corona, reduced surface filling factors due to polar concentration of magnetic flux, and changes in dynamo efficiency at extreme rotation rates; however, its physical origin remains unresolved.

In \citet{doyle2019}, we performed a statistical analysis of stellar flares in a sample of 149 low-mass dwarfs (M0 -- M6 V) using 2-min cadence {\tess} light curves from Sectors 1 -- 3. From this sample, we identified nine ultra-fast rotating (UFR) low-mass stars with rotation periods $P_{\rm rot} \leq 0.3$~d that exhibit unexpectedly low levels of flaring activity. We found no evidence that this suppressed activity is related to stellar age or projected rotational velocity. This finding was also supported by \citet{Gunther2020}, who used data taken from the first two months of the {\tess} mission, and found a `tentative' decrease in the flare rate for stars with P$<$0.3 d. This result is surprising in the context of the established rotation–activity relation, which predicts increasing activity with faster rotation \citep{hartmann1987rotation}. Extending this work, \citet{ramsay2020ufr} analysed {\tess} 2-min cadence data from Sectors 1 -- 13 and identified over 600 rapidly rotating low-mass stars (K9 -- M5 V) with $P_{\rm rot} < $1~d. They showed the fraction of stars exhibiting flares declines sharply for rotation periods shorter than $P_{\rm rot} < 0.2$~d compared to stars with 0.6 $< P_{\rm rot} < 1.0$~d, reinforcing the presence of reduced activity at extreme rotation rates. It is known that flares emit more prominently at bluer wavelengths; therefore, it is possible that these studies missed many shorter duration lower energy flares. 

In a detailed analysis of the UFR (P$_{\rm{rot}}$ = 0.117~d) M dwarf LP 89–187, \citet{ramsay2025hipercam} used HiperCam observations to monitor for optical flares at lower energy but failed to detect any flares over the short observation span. Furthermore, the flare energy distribution of LP 89-187 derived from {\tess} data indicated it has a lower flare rate than stars in the $\beta$ Pic moving group, which have an age of $\sim$24 Myr.

In \citet[][hereafter Paper I]{doyle2022}, we obtained spectropolarimetric observations of ten UFRs ($P_{\rm rot} < 1$~d) using VLT/FORS2 and detected line-of-sight magnetic fields of order 1 -- 2~kG in five targets. However, given that only half of the sample showed detectable fields, and four of these were among the more active stars, it appears unlikely that magnetic field strength alone can account for the suppressed flaring activity observed in UFRs. Therefore, in this paper, we investigate the X-ray luminosity of UFR low-mass stars to explore the possibility of supersaturation. In \S \ref{sec:sample} we define our sample of low-mass UFRs, providing details on stellar interiors and stellar multiplicity. All of our observational data, both X-ray and optical, are detailed in \S \ref{sec:obs}, which is followed by \S \ref{sec:x-ray_analysis} and \ref{sec:tess_analysis} explaining our analysis of each set of observations from stellar flare identification to extracting the X-ray luminosities. Finally, in \S \ref{sec:Discussion} we bring together all our findings and discuss the implications for our UFR low-mass sample. 

\begin{figure}
    \centering
    \includegraphics[width=0.99\linewidth]{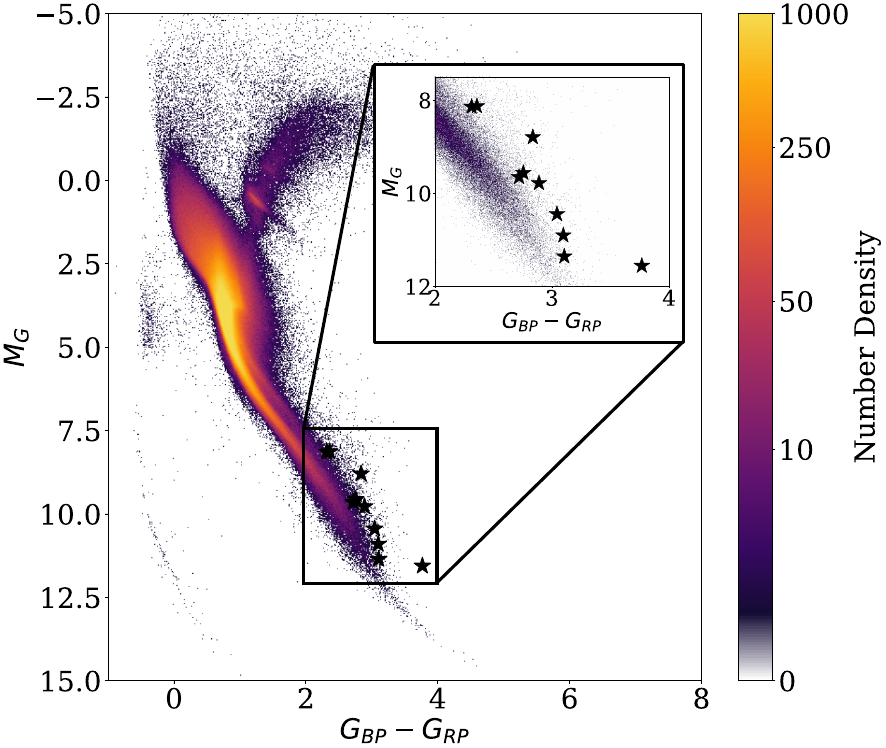}
    \caption{A Hertzsprung-Russell (HR) diagram of the {\tess}-SPOC FFI sample described in \citet{doyle2024}, generated using {\sl Gaia} DR3 \citep{vallenari2023gaiadr3} colours and parallax, where the colour bar represents the log density of stars. Our ten UFR low-mass stars are overplotted as black star markers to show their location. Note: the parallax for TIC 141807839 was taken from {\sl Gaia} DR2, but all other colours and parallaxes were taken from DR3 for our sample. }
    \label{fig:HRD}
\end{figure}

\section{Low-mass star sample}
\label{sec:sample}
We utilise the same low-mass star sample as in Paper I, which comprises ten stars in the period range 0.1 -- 1.0~d. These targets are split into five equal bins, each with two stars -- one showing relatively few flares with {\tess} and the other showing an increase in flare activity. All of our targets are brighter than $T_{\rm{mag}}$ = 13.2 with spectral types between M2 and M5.5. For full details on the stellar properties of each target, we refer the reader to \citet{doyle2022}; however, in this paper, we summarise these in Table \ref{stellar_prop} for completeness. The quiescent luminosity of each of our stars was determined in \citet{doyle2019}. The quiescent flux was estimated for each star by convolving a template spectrum, constructed from SkyMapper multicolour magnitudes converted to flux, with the {\tess} band-pass. Inverted {\sl Gaia} parallaxes were used to provide distances to each star, which were combined with the quiescent flux to determine the quiescent stellar luminosity.

In Figure \ref{fig:HRD} we show the location of all ten of our UFR low-mass stars on the colour-magnitude diagram created using {\sl Gaia} DR3 \citep{vallenari2023gaiadr3}. It can be observed that our targets lie slightly above the main sequence branch, indicating stellar multiplicity and/or that they are still pre-main sequence. Therefore, we stress that our sample is not indicative of a typical sample of field M dwarfs, but has been selected to be consistent with our other studies. Multiplicity is indeed observed or heavily implied for some of these systems as we discuss in \S \ref{ssec:multiplicity}, while the latter may also be expected for some of the least massive stars in the sample, which can take in excess of 100~Myr to reach the main sequence \citep[e.g.][]{Palla1999}.

By the late pre-main sequence, the difference in the expected level of X-ray emission compared to young main-sequence stars is rather subtle. Indeed our comparison sample in \S \ref{sec:Discussion}, that of \citet{Wright2011,Wright2018}, likely includes many late pre-main sequence stars at the low mass end, and yet there is no deviation away from $L_{\rm X}/L_{\rm bol} \sim 10^{-3}$ at the saturated level of low mass stars in their sample. While their work does describe a cut of any star below 10\,Myr to exclude pre-main sequence stars, many low mass stars remain in the pre-main sequence phase well past this age, as mentioned above, and would not have been cut. All of our stars with known ages are older than 10\,Myr (Table \ref{stellar_prop}), in line with the Wright et al. sample. Moreover, other past studies have shown that stars below 1\,M$_\odot$ cluster near $10^{-3}$ at even younger ages of just a few Myr \citep[e.g.][]{Preibisch2005,Getman2022}, well within the pre-main sequence epoch. However, we also note \citet{Argiroffi2016} found that at 13~Myr (the approximate lower end of the age range among our sample) stars in the 0.3 -- 0.7~M$_\odot$ range may be slightly higher in $L_{\rm X}/L_{\rm bol}$ than their higher mass or young main sequence counterparts.

\begin{table*}
\caption{The stellar properties of the ten low-mass stars in our sample: we show the catalogue name; TIC ID \citep{stassun2018tess}; RA and DEC; stellar masses and radii from the TIC \citep{stassun2018tess}; MV Spectral type ({\tt SIMBAD}); the stellar effective temperature; $T_{mag}$ \citep{stassun2018tess}; stellar age; stellar rotation period (this work); RUWE and Distance from {\sl Gaia} DR3 \citep{vallenari2023gaiadr3} - note target AL 442 taken from {\sl Gaia} DR2 \citep{brown2018gaiadr2} - and the quiescent luminosity \citep{doyle2019}. Stellar ages are taken from the following sources using estimates from moving groups and associations: $^a$ \citet{zhan2019complex}, $^b$ \citet{janson2017binaries}, $^c$ \citet{loyd2021hazmat}, $^d$ \citet{booth2021age} and $^e$ \citet{gagne2014banyan}.}

   \begin{center}
   \label{stellar_prop}
\resizebox{1.0\textwidth}{!}{
	\begin{tabular}{llcccccccccccc}
    \hline 
	Name                & TIC ID    & RA      & Dec       & Mass           & Radius      & SpT  & T$_{\rm{eff}}$ & T$_{\rm{mag}}$ & Age   &  P$_{\rm{rot}}$ & RUWE  & Distance  & $\log(\rm{L}_{\rm{star}})$  \\
	                      &           & (J2000) & (J2000)   & M$_{\odot}$    & R$_{\odot}$ &      & (K)            &                & (Myr) &  (days)         &       &           &                             \\
    \hline
    UCAC4 204-001345    & 158596311 & 22.1269 & -49.3528  & 0.36 & 0.37 & 4.1  & 3159 &  12.3  & $\sim$45$^b$  & 0.15 & 1.45 &  43.31    &  31.14   \\
    UCAC3 53-724        & 425937691 & 5.3666  & -63.8525  & 0.31 & 0.32 & 5.5  & 2877 &  13.2  & $\sim$45$^a$  & 0.10 & 2.85 &  43.14    &  30.80   \\
    UPM J0113-5939      & 206544316 & 18.4189 & -59.6598  & 0.46 & 0.46 & 3.7  & 3237 &  11.6  & $\sim$150$^a$ & 0.32 & 1.26 &  42.85    &  31.39   \\ 
    2MASS J0033-5116    & 156002545 & 8.3524  & -51.2790  & 0.45 & 0.46 & 3.4  & 3318 &  11.4  & $\sim$45$^b$  & 0.35 & 1.51 &  41.34    &  31.42   \\
    GSC 04683-02117     & 248354845 & 17.8474 & -5.4273   & 0.47 & 0.48 & 3.5  & 3300 &  11.1  & --            & 0.52 & 1.64 &  36.32    &  31.47   \\ 
    2MASS J0146-5339    & 229142295 & 26.6237 & -53.6596  & 0.24 & 0.27 & 4.5  & 3131 &  11.2  & --            & 0.45 & 3.32 &  17.40    &  30.79   \\ 
    GSC 08859-00633     & 220539110 & 43.4470 & -61.5878  & 0.65 & 0.68 & 3.0  & 3538 &  9.9   & --            & 0.77 & 7.52 &  40.08    &  31.98   \\ 
    EXO 0235.2-5216     & 166808151 & 39.2163 & -52.0510  & 0.66 & 0.69 & 2.0  & 3510 &  9.9   & $\sim$45$^c$  & 0.74 & 1.26 &  38.83    &  31.95   \\
    AL 442              & 141807839 & 92.8752 & -72.2271  & 0.61 & 0.63 & 4.5  & 3259 &  11.2  & $\sim$13$^d$  & 0.85 & --   &  56.93    &  31.81   \\
    2MASS J0232-5746    & 201861769 & 38.0806 & -57.7699  & 0.30 & 0.32 & 4.1  & 3134 &  12.8  & $\sim$45$^e$  & 0.87 & 1.29 &  45.84    &  30.97   \\
    \hline
    \end{tabular}}
    \end{center}
\end{table*}

\subsection{Stellar Interiors}
A dynamo mechanism in stellar interiors generates the magnetic field of a star. This, coupled with convective motions and stellar rotation, drives magnetic activity, such as stellar flares. Through equations of stellar structure and evolution, \cite{chabrier1997structure} explores the fully convective limit of low-mass stars, finding that the minimum mass for the onset of radiation in the core is 0.35~M$_{\odot}$. Within our low-mass stellar sample, three of our stars (UCAC3 53-724, 2MASS J0146-5339 and 2MASS J0232-5746) are below this threshold, and one (UCAC4 204-001345) is on the cusp of the threshold. As a result, for our analysis and discussions, we consider these four targets to possess fully convective interiors, though we note that stars slightly above this limit may remain fully convective if they are still pre-main-sequence. However, while fully convective stars are considered to generate their magnetic field and corona through a different dynamo mechanism, X-ray studies to date have shown remarkably similar evolutionary behaviour to partially convective stars \citep{Wright2016,Wright2018}.

\subsection{System Multiplicity}
\label{ssec:multiplicity}
The existence of companion stars in our systems, particularly unresolved companions, may impact our results. We noted in Paper I the wide use of \textit{Gaia} Renormalised Unit Weight Error (RUWE) values as a possible indicator of multiplicity, as a higher RUWE indicates a poorer astrometric fit which may be driven by the binary motion of an unresolved multiple stellar system. Various thresholds for binarity have been used in the literature, such as 1.4 \citep[e.g.][]{Lindegren2021} and 1.25 \citep{Penoyre2022}, though even some stars well below RUWE~=~1.25 have been demonstrated to be binaries \citep[e.g.][]{OBrien2025}. We list RUWE values from DR3 in Table~\ref{stellar_prop}, with all of these being above the lower 1.25 threshold, half being above 1.4, and three being well in excess of 2. The highest value (RUWE~=~7.52) is for GSC 08859-00633, for which radial velocity (RV) observations in Paper I showed a clear modulation due to a likely brown dwarf companion. That RV investigation also revealed possible line splitting in some of the spectra for GSC 04683-02117.

Alongside this, several of our other systems have existing evidence of (or discussion of the lack thereof) multiplicity in the literature. We summarise this for each target in turn. UCAC3 53-724 was identified as a wide quadruple system by \citet{Gonzalez-Payo2024}, although each of the other components is over 30\,arcmin away, and thus none are blended with our target in any of our observations. 2MASS J0146-5339 (WT 50) is suggested by \citet{Bertini2023} as being a secondary companion to *q01 Eri, although at almost 36\,arcmin away on the sky, the primary does not affect our study. \citet{Janson2017} determines UCAC4 204-0011345 and 2MASS J0033-5116 to be single stars down to detection limits. \citet{Zuniga2021} determines 2MASS J0232-5746 to also show no evidence of multiplicity, while identifying EXO 0235.2-5216 as a possible but unconfirmed spectroscopic binary. UPM J0113-5939 has been determined to be an eclipsing binary in {\tess} \citep{Prsa2022}. AL 442 has been spatially resolved to be at less than 0.2\,arcsec from an M4+M5 pair, which are likely, but as yet not confirmed, to be physically bound \citep{Janson2012,Calissendorff2022}. Neither \textit{Swift} nor {\tess} can separate the two components for any close binaries among the sample, and thus our flux measurements for such systems represent the total emission from all component stars.

\section{Observations}
\label{sec:obs}
In this study, we utilise both X-ray observations from {\sl Swift} and {\sl XMM-Newton}, along with photometric observations from {\tess}. In this section, each data source is detailed with a summary provided in Table \ref{tab:observations}. 

\begin{table}
    \centering
    \caption{Summary of the data used in this work.}

    {\sl {\sl Swift} Data}
    \begin{tabular}{lcclc}
    \hline
    \hline
    TIC  & \textit{Swift} & ObsIDs  & Date Span         & XRT PC  \\
    ID   & Target        &         &                   & exp. time     \\
         & ID            &          &                   & (ks)     \\
    \hline
    158596311 & 3112524 & 003-039 & 8 May 24 - 19 Jan 25  & 28.4         \\
    425937691 & 3112523 & 001-121 & 30 Mar 23 - 14 Feb 24 & 84.9         \\
    206544316 & 3112525 & 001-032 & 30 Aug 23 - 30 Nov 24 & 11.0         \\
    156002545 & 3112526 & 001-132 & 19 May 23 - 8 Jan 25  & 78.2         \\
    248354845 & 3112527 & 001-039 & 22 Mar 23 - 5 Jul 23  & 81.0         \\
    229142295 & 3112528 & 001-143 & 16 Mar 23 - 14 May 24 & 88.9         \\
    220539110 & 3112529 & 001-137 & 17 Mar 23 - 8 Jan 25  & 77.4         \\
    166808151 & 3112530 & 001-145 & 18 Mar 23 - 13 May 24 & 100.0        \\
    141807839 & 3112531 & 001-146 & 29 Apr 23 - 6 Jan 25  & 70.0         \\
    201861769 & 3112532 & 001-254 & 19 Mar 23 - 14 May 24 & 115.3        \\ \hline
    \end{tabular}
    \vspace{5mm}

    {\sl XMM-Newton}
    \begin{tabular}{lccc}
    \hline
    \hline
    TIC ID  & ObsIDs  & Date         & pn exp. time  \\
     & & & (ks) \\
    \hline
    201861769 & 0743070501 & 31 May 2014  & 12.5 \\
    \hline
    \end{tabular}
    \vspace{5mm}
    
    {\sl Photometric {\tess} Data}
    \begin{tabular}{llccc}
    \hline 
    \hline 
    TIC ID & Sectors & Obs. length & $t_{\rm{exp}}$  \\
           &         & (days)            & (s)             \\
    \hline
    158596311 &  69 \& 96                        & 40.69  &  120 \\
    425937691 &  68, 69, 95 \& 96                & 84.06  &  120 \\
    206544316 &  68, 69, 95 \& 96                & 84.03  &  120 \\
    156002545 &  69 \& 96                        & 41.25  &  120 \\
    248354845 &  70$^*$                          & 20.76  &  200 \\
    229142295 &  69 \& 96                        & 41.94  &  120 \\
    220539110 &  68$^*$ \& 69$^*$                & 40.28  &  200 \\
    166808151 &  69 \& 96                        & 41.75  &  120 \\
    141807839 &  61, 63-69, 87, 88, 90 \& 93-96  & 314.66 &  120 \\
    201861769 &  69 \& 96                        & 41.46  &  120 \\ 
    \hline
    \end{tabular}
    
        \vspace{2mm}
     \begin{flushleft}
    {\bf Notes:} For {\sl Swift} data, not all ObsIDs between the start and end value exist in the archive, but all observations within the date range were used in our analysis. $^*$ For these {\tess} Sectors 200~s FFI light curves processed by SPOC were used as no 2-minute cadence light curves were available. 
    \end{flushleft}
    \label{tab:observations}
\end{table}

\subsection{{\sl Swift}}
\textit{The Neil Gehrels Swift Observatory} \citep{Gehrels2004} has operated in low Earth orbit since 2004, with a primary function of detecting and characterising gamma-ray bursts. However, the multi-wavelength span of its three instruments (gamma-ray, soft X-ray, and ultraviolet/optical) has enabled a wide range of science.

We observed all ten stars in our sample with \textit{Swift} at multiple epochs between March 2023 and January 2025. Table \ref{tab:observations} shows an overview of these observations. Our analyses were focused on the soft X-ray, using the data taken with the X-Ray Telescope \citep[XRT;][]{Burrows2004} instrument in Photon Counting (PC) mode, which operates from 0.3--10\,keV. We used a mix of manual data reduction processes and the online tools for building XRT products\footnote{\url{https://www.swift.ac.uk/user_objects/}} provided by the UK Swift Science Data Centre \citep[UKSSDC;][]{Evans2007,Evans2009}. For the manual reductions, we used \textsc{HEASoft}\footnote{\url{https://heasarc.gsfc.nasa.gov/docs/software/lheasoft/}} version 6.33.2.

\subsection{{\sl XMM-Newton}}
\textit{XMM-Newton} \citep{Jansen2001} is the European Space Agency's flagship X-ray observatory, providing a slew of imaging and spectroscopic instruments across its three X-ray telescopes, as well as simultaneous optical/ultraviolet observations with the Optical Monitor. \textit{XMM-Newton} possesses a larger effective area than \textit{Swift}, while operating in a similar energy band (0.2--12\,keV). Its high eccentricity orbit means it is also capable of continuous stares up to $\sim$35~hrs long, in contrast to \textit{Swift's} short snapshots, which often require stacking.

One of the stars in our sample, 2MASS J0232-5746, was observed by \textit{XMM-Newton} on 2014 May 31 (ObsID: 0743070501; PI: Principe). We analysed this archival observation to complement our \textit{Swift} analysis, and compare the brightness of the star a decade apart to search for longer-term variation. Our analysis focused on the data taken with the three detectors (pn, MOS1, and MOS2) comprising the European Photon Imaging Cameras \citep[EPIC;][]{Strueder2001,Turner2001}. All three detectors were used in full frame mode with medium optical blocking filters. 

\subsection{\tess}
The Transiting Exoplanet Survey Satellite \citep[{\tess}:][]{Ricker2015} was launched in April 2018 with a primary mission of detecting small exoplanets via the transit method around bright nearby stars. Its primary 2-year mission was to complete a near ($\sim$75\%) all-sky survey observing more than 200,000 stars within the solar neighbourhood. Since then, {\tess} has completed two further all-sky surveys and covered the ecliptic plane during three extended missions and is currently in its eighth year of observations with a ninth planned. In \cite{doyle2022}, we analysed the flare properties of the ten stars in this sample utilising {\tess} light curves from Cycles 1 and 3 (Sectors 1--13 \& 27--39, respectively). In this study, we use light curves from both Cycles 5 \& 6 (Sectors 56--70) and Cycle 7 (Sectors 84--96), computing a similar analysis but focusing on the long-term flare rates and stellar variability.  

The {\tess} photometric light curves used in this analysis were observed between 2022 Sep 01 and 2023 Oct 16 for Sectors 56--70 and 2024 Oct 01 and 2025 Sep 15 for Sectors 84--96. At the time of writing, data up to Sector 99 was available; however, this includes all data in 2-minute cadence for our targets in Cycle 5 and 7. We downloaded the calibrated 2-minute light curves for each of our target stars from the Mikulski Archive for Space Telescopes (MAST) data archive\footnote{\url{https://archive.stsci.edu/tess/}}, using the data values for {\tt PDCSAP\_FLUX}. All points which did not have {\tt QUALITY=0} were removed, and each light curve was normalised by dividing the flux of each point by the mean flux of the star \citep[see][for more details]{doyle2019,ramsay2020ufr}. 

Two of our targets, GSC 04683-02117 (TIC 248354845) and GSC 08859-00633 (TIC 220539110), did not have 2-minute cadence light curves available in Cycles 5 and 7. Therefore, for these targets, we use the Science Processing Operations Centre \citep[SPOC: ][]{jenkins2016tessspoc} Full Frame Image (FFI) 200-second light curves. From Sector 57 onwards, the FFI cadence was improved from 10-min to 200~s, making this mode comparable to the 2-minute postage stamp targeted sample. SPOC began processing FFI light curves for up to 160,000 targets per Sector \citep[see][for full details]{caldwell2020spoctargets,doyle2024} during the second year of {\tess} observations. All pixel and light curve data from the SPOC are contributed as High-Level Science Products on MAST, where they are publicly available. At the time of writing, SPOC has processed all FFI targets up to and including Sector 80; therefore, we do not have Cycle 7 data for GSC 04683-02117 and GSC 08859-00633. In Table \ref{stellar_prop}, we list the {\tess} Sectors which are used in this paper for each target and identify where 200~s FFI light curves are used.

\section{X-ray analysis and results}
\label{sec:x-ray_analysis}

We focused our X-ray analyses on energies below 2.4\,keV. We made this choice to allow for easy comparison with previous population studies, many of which derive from \textit{ROSAT}, for which catalogued values are typically given for the 0.1--2.4\,keV band.

We verified that each star was successfully detected by combining the event files for each star separately in \textsc{Xselect} 2.5b and extracting images. For all ten images, a source was visually apparent at the expected location of the star.

\subsection{X-ray light curves}

\begin{figure*}
\centering
 \includegraphics[width=\textwidth]{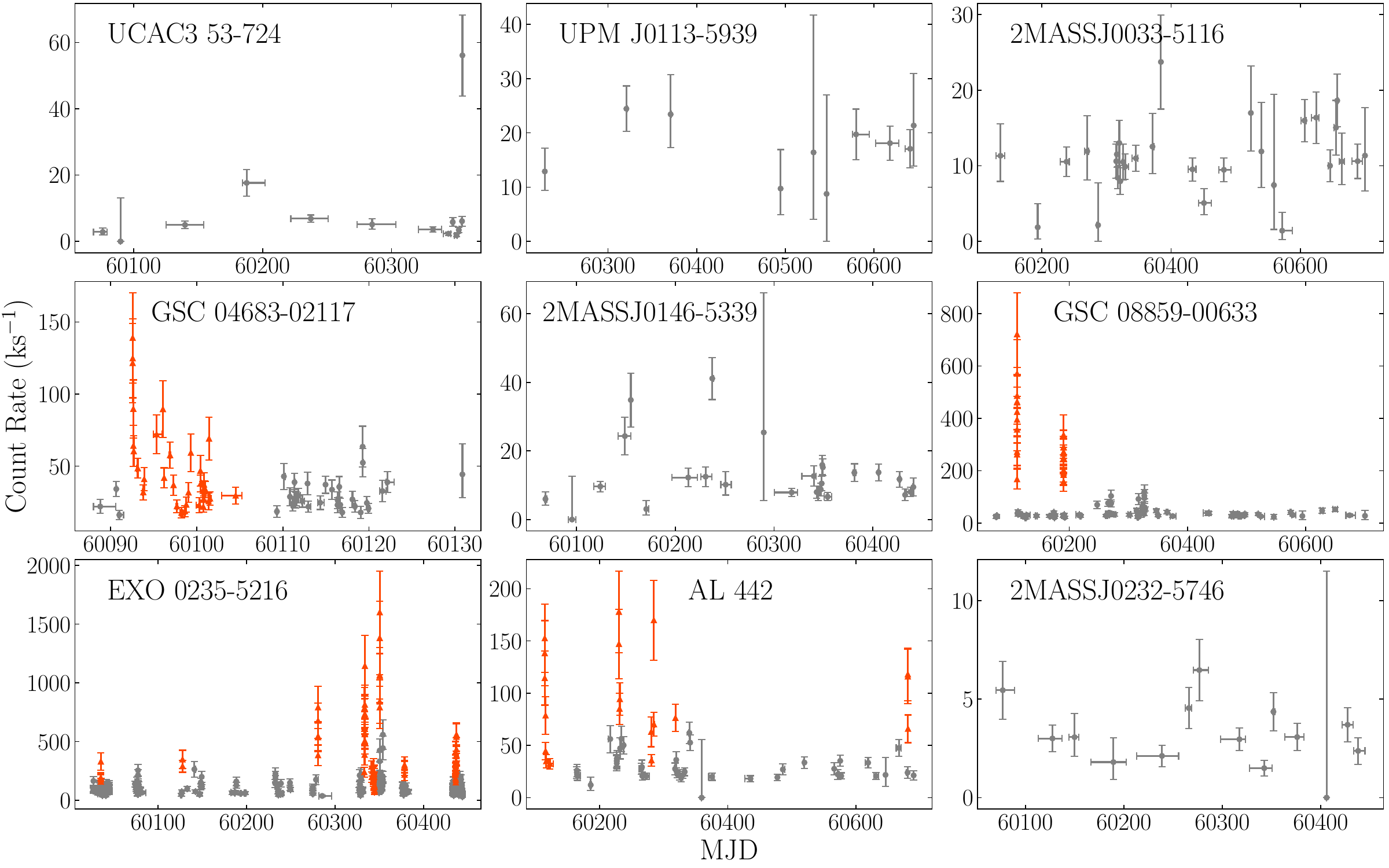}
 \caption{\textit{Swift} XRT light curves for nine of our ten stars in the 0.3--2.4\,keV band, with a minimum of 20 counts per bin. Points denoted as ``quiescent" are displayed with grey circles, while ``elevated/flaring" epochs are denoted with red triangles.}
 \label{fig:xLC}
\end{figure*}

To build light curves in the 0.3 -- 2.4\,keV energy band, we used the UKSSDC online tool \citep{Evans2007,Evans2009}, with the background-subtracted results for nine of the ten stars plotted in Figure~\ref{fig:xLC}. We do not display a light curve for UCAC4\,204-001345 because the data did not contain enough counts to build a light curve that is informative beyond ruling out strong flaring. We forced the minimum counts per bin to be 20 wherever possible, turned dynamic binning off, and set the ``maximum orbital gap" option to one day. The tool calculates errors for bins with more than 15 counts using Gaussian statistics, and Bayesian statistics for those with less, as per \citet{Evans2007,Evans2009} Any bins that were upper limits resulting from a lack of signal were discarded from the light curve plot, though they were included in the spectra extractions in Section~\ref{ssec:xSpectra}.

We searched for temporal variations in the X-ray emission of each star by eye, identifying four stars with substantial visible changes in their respective light curves: GSC 04683-02117, GSC 08859-00633, EXO 0235.2-5216, and AL 442. A few others showed hints of variation, but either the very short duration of the increase or the measurement uncertainties on the higher points were such that we deemed them constant for the purposes of extracting spectra in Section~\ref{ssec:xSpectra}. Many of the epochs identified are likely X-ray flares with a characteristic sudden increase followed by an exponential decay. A notable exception is GSC 04683-02117, which shows a complex evolution of the X-ray emission between MJD 60090 and 60105. There are a few clear flares within this epoch, but it appears to be imprinted upon some slightly longer-term evolution over a few weeks, possibly resulting from a particularly bright active region enhancing the baseline emission. There are a few other epochs for EXO 0235-5216 and AL 442 where flare-like behaviour is observed to decay over a few days rather than the typical timescale of hours. As a result, we use the term ``elevated" rather than ``flaring" to collectively refer to all of the epochs with increased emission above surrounding points. These are highlighted in red in Figure~\ref{fig:xLC}, with all other epochs considered quiescent.

The \textit{XMM-Newton} light curve for 2MASS J0232-5746 showed a hint of a decrease at harder energies ($\gtrsim0.85$\,keV) over the first half of the observation. This could potentially be the tail end of a flaring event, which peaked prior to the beginning of the observation. However, the uncertainties on the points were large enough and the slope shallow enough that we chose not to separate this into a separate elevated epoch.

\subsection{Spectra}
\label{ssec:xSpectra}

For the six stars we identified as non-varying, we reduced the data manually, again combining event files in \textsc{Xselect}. We used 20 pixel (47.1\,arcsec) radius source extraction regions for all stars, except 2MASS J0232-5746, for which we used a slightly smaller 16 pixel (37.7\,arcsec) radius due to a nearby, unrelated source. Our background regions were set to 60 pixel radii, except UCAC3 53-724, for which we used a radius of 50 pixels due to several nearby sources. We made custom Auxiliary Response Files (ARFs) by following the processes outlined on the UKSSDC website\footnote{\url{https://www.swift.ac.uk/analysis/xrt/arfs.php}}.

For the four stars with at least one identified epoch of flaring or elevated count rate, we visually determined the start and end of each such epoch. We then extracted separate spectra for each, together with a single spectrum for all other data for that star, which we designated as quiescent. Splitting up the data in this way significantly complicated the process for making custom ARFs, and so we reverted to using the UKSSDC online product builder tool to extract spectra, background spectra, and ARFs for these four stars.

We fitted our spectra using \textsc{xspec} 12.14.0h \citep{Arnaud1996}, using between one and three APEC models \citep{Smith2001} combined additively. For those stars with elevated epochs, we fitted each extracted spectrum separately. We fixed the abundances to the Solar values provided by \citet{Asplund2009}, except for 2MASS J0146-5339, for which we could only obtain a good fit to the spectrum by freeing up the abundances using a single scaling factor to Solar. This factor was best fit to $0.21^{+0.10}_{-0.07}$.

The contribution from interstellar absorption is small for all of our targets as they are all nearby, and because the young age of these stars leads them to have hotter plasma and thus harder spectra than older, field age stars \citep[e.g.][]{Johnstone2021}. However, for completeness, we included a TBABS term to account for this absorption \citep{Wilms2000}, fixing the neutral hydrogen column densities to values we calculated using the \textsc{lism} code\footnote{\url{https://github.com/allisony/LISM_NHI}} \citep{Youngblood2025}. In all cases, our neutral hydrogen column densities were $< 5.1\times10^{18}$\,cm$^{-2}$, and the effect on our fluxes was less than 0.25 per cent - well within our measurement uncertainties. We plot the quiescent XRT spectrum for each star in Figure~\ref{fig:swiftSpec}, together with their best-fitting models.

For the \textit{XMM-Newton} observation of 2MASS J0232-5746, all data were treated as quiescent. We extracted spectra separately for each of the three EPIC cameras using a 20\,arcsec radius source region, and several larger background regions up to 60\,arcsec in radius. We performed a single joint fit across the spectra from each camera, using identical \textsc{xspec} fitting procedures as used for the \textit{Swift} data. Our \textit{XMM-Newton} spectra for this star are plotted in Figure~\ref{fig:XMMspec}, with the best fit model overlaid.

\subsection{Fluxes and luminosities}
\label{ssec:fluxes}

\begin{table*}
\centering
\caption{Best fit parameters for the fits to each star's combined epochs defined as quiescent. We quote each APEC temperature and associated emission measure. We give the unabsorbed fluxes at Earth for each star in two bands: the directly observed 0.3--2.4\,keV band (labelled with $_{\rm 0.3}$), and the commonly used ROSAT 0.1--2.4\,keV band (labelled with $_{\rm 0.1}$). We also provide X-ray luminosities and the ratio of this to the bolometric luminosity for the latter band.}
\label{tab:temps+fluxes}
\begin{threeparttable}
\resizebox{1.0\textwidth}{!}{
\begin{tabular}{lcccccccccc}
\hline
TIC ID        & $kT_1$                    & $EM_1$                 & $kT_2$                    & $EM_2$                & $kT_3$                 & $EM_3$                & $F_{\rm X,0.3}$         & $F_{\rm X,0.1}$        & $L_{\rm X,0.1}$           & $\frac{L_{\rm X,0.1}}{L_{\rm bol}}$ \\
              & (keV)                     & ($^a$)                 & (keV)                     & ($^a$)                & (keV)                  & ($^a$)                & ($^b$)                  & ($^b$)                 & ($^c$)                    &    ($^d$)                    \\ \hline
158596311     & $1.79^{+1.09}_{-0.40}$    & $7.5^{+2.3}_{-2.1}$    & -                         & -                     & -                      & -                     & $3.58^{+0.96}_{-0.54}$  & $4.25^{+1.22}_{-0.64}$ & $0.95^{+0.27}_{-0.14}$    & $0.69^{+0.20}_{-0.10}$      \\[0.1cm]
425937691     & $0.794^{+0.084}_{-0.096}$ & $3.39^{+0.81}_{-0.71}$ & $1.94^{+0.76}_{-0.34}$    & $9.6^{+2.0}_{-1.9}$   & -                      & -                     & $7.97^{+1.09}_{-0.39}$  & $9.52^{+1.21}_{-0.52}$ & $2.12^{+0.27}_{-0.12}$    & $3.36^{+0.43}_{-0.19}$      \\[0.1cm]
206544316     & $0.247^{+0.054}_{-0.049}$ & $19.5^{+5.1}_{-4.6}$   & $1.189^{+0.099}_{-0.091}$ & $37.5^{+5.7}_{-4.9}$  & -                      & -                     & $40.5^{+3.1}_{-3.6}$    & $51.5^{+5.0}_{-5.3}$   & $11.3^{+1.1}_{-1.2}$      & $4.61^{+0.45}_{-0.47}$      \\[0.1cm]
156002545     & $0.320^{+0.027}_{-0.022}$ & $15.8\pm1.5$           & $1.187^{+0.101}_{-0.086}$ & $12.0^{+1.5}_{-1.3}$  & -                      & -                     & $22.01^{+0.96}_{-0.90}$ & $27.3^{+1.4}_{-1.3}$   & $5.58^{+0.28}_{-0.26}$    & $2.12^{+0.11}_{-0.10}$      \\[0.1cm]
248354845     & $0.286\pm0.014$ & $32.4^{+2.3}_{-2.2}$   & $1.57^{+0.13}_{-0.11}$    & $41.3^{+3.7}_{-3.6}$  & -                      & -                     & $62.1^{+2.1}_{-1.8}$    & $77.6^{+3.0}_{-2.6}$   & $12.25^{+0.47}_{-0.41}$   & $4.15^{+0.16}_{-0.14}$      \\[0.1cm]
229142295$^e$ & $0.283^{+0.048}_{-0.034}$ & $7.3^{+1.4}_{-1.2}$    & $1.052^{+0.89}_{-0.085}$  & $6.2^{+1.5}_{-1.2}$   & -                      & -                     & $22.53^{+0.50}_{-0.21}$ & $31.5^{+1.3}_{-2.8}$   & $1.142^{+0.049}_{-0.100}$ & $1.852^{+0.079}_{-0.162}$   \\[0.1cm]
220539110     & $0.305^{+0.029}_{-0.034}$ & $28.2^{+3.9}_{-4.5}$   & $0.886^{+0.086}_{-0.077}$ & $21.2\pm3.3$          & $2.21^{+0.46}_{-0.26}$ & $71.0^{+5.5}_{-5.6}$  & $83.7^{+2.8}_{-1.1}$    & $102.3^{+4.0}_{-1.6}$  & $19.67^{+0.80}_{-0.37}$   & $2.059^{+0.084}_{-0.039}$   \\[0.1cm]
166808151     & $0.270^{+0.011}_{-0.012}$ & $77.9^{+4.5}_{-5.0}$   & $1.023^{+0.086}_{-0.077}$ & $64.3^{+5.9}_{-5.0}$  & $5.4^{+2.2}_{-1.2}$    & $165.0^{+9.2}_{-9.5}$ & $210.5^{+3.6}_{-2.3}$   & $260.6^{+5.5}_{-3.3}$  & $47.04^{+0.99}_{-0.63}$   & $5.277^{+0.111}_{-0.070}$   \\[0.1cm]
141807839     & $0.295\pm0.021$           & $57.0^{+9.6}_{-7.0}$   & $1.06^{+0.13}_{-0.10}$    & $36.1^{+13.2}_{-7.9}$ & $5.4^f$                & $89^{+14}_{-15}$      & $60.5^{+2.7}_{-2.6}$    & $74.8^{+3.9}_{-4.4}$   & $29.0^{+2.1}_{-2.2}$      & $4.49^{+0.32}_{-0.33}$      \\[0.1cm]
201861769     & $0.145^{+0.029}_{-0.020}$ & $8.1^{+3.0}_{-2.2}$    & $0.779^{+0.052}_{-0.051}$ & $4.15\pm0.38$         & -                      & -                     & $6.55^{+0.23}_{-0.89}$  & $12.7^{+2.6}_{-3.4}$   & $3.20^{+0.66}_{-0.86}$    & $3.43^{+0.71}_{-0.93}$      \\ \hline
\end{tabular}
}
\begin{tablenotes}
\item $^a$ $10^{50}$\,cm$^{-3}$
\item $^b$ $10^{-14}$\,erg\,s$^{-1}$\,cm$^{-2}$
\item $^c$ $10^{28}$\,erg\,s$^{-1}$
\item $^d$ $10^{-3}$ (dimensionless)
\item $^e$ The fit for TIC\,229144295 (2MASS J0146-5339) included an extra free parameter for elemental abundances, which best fit to $0.211^{+0.098}_{-0.068}$ times Solar. All \newline other stars were fixed at Solar abundances \citep{Asplund2009}.
\item $^f$ The highest temperature for TIC\,141807839 was unconstrained at the upper end, running up to the hard limit of 64\,keV. The lower 1-$\sigma$ confidence interval \newline was 3.2\,keV.
\end{tablenotes}
\end{threeparttable}
\end{table*}

\begin{table*}
\centering
\caption{Comparison of the best fit quiescent fluxes to the various defined elevated epoch fluxes of the same stars. All fluxes are in units of $10^{-14}$\,erg\,s$^{-1}$\,cm$^{-2}$ for the observed 0.3--2.4\,keV band.}
\label{tab:elevatedFluxes}
\begin{tabular}{lcccccccc}
\hline
TIC ID    & Quiescent             & Elevated 1            & Elevated 2           & Elevated 3           & Elevated 4        & Elevated 5           & Elevated 6        & Elevated 7        \\ \hline
248354845 & $62.1^{+2.1}_{-1.8}$  & $113.4^{+9.4}_{-6.0}$ & $63.6^{+5.2}_{-4.0}$ & $66.5^{+4.3}_{-3.4}$ & -                 & -                    & -                 & -                 \\[0.1cm]
220539110 & $83.7^{+2.8}_{-1.1}$  & $834^{+72}_{-48}$     & $470^{+35}_{-36}$    & -                    & -                 & -                    & -                 & -                 \\[0.1cm]
166808151 & $210.5^{+3.6}_{-2.3}$ & $490^{+47}_{-51}$     & $627^{+107}_{-66}$   & $1410^{+250}_{-230}$ & $522^{+22}_{-18}$ & $2360^{+240}_{-230}$ & $560^{+97}_{-51}$ & $599^{+59}_{-50}$ \\[0.1cm]
141807839 & $60.5^{+2.7}_{-2.6}$  & $108^{+14}_{-11}$     & $234^{+39}_{-15}$    & $123^{+11}_{-14}$    & $160^{+45}_{-18}$ & $195^{+22}_{-156}$   & -                 & -                 \\ \hline
\end{tabular}
\end{table*}

In Table~\ref{tab:temps+fluxes}, we show the best fitting model parameters for the quiescent \textit{Swift} XRT spectra of each star, along with fluxes in two slightly different energy bands. In both cases, the 2.4\,keV upper bound is based on the \textit{ROSAT} band for direct comparison with the numerous population studies based on its results. The flux of these stars above those energies with XRT is also small. $F_{\rm X,0.3}$ is the 0.3 -- 2.4\,keV flux, corresponding to the directly observable band of these stars with XRT, which lacks sensitivity below 0.3\,keV. $F_{\rm X,0.1}$ is the 0.1 -- 2.4\,keV flux, matching the \textit{ROSAT} band itself, and therefore includes a contribution from the unobserved 0.1 -- 0.3\,keV energy range based on an extrapolation of our best fitting models into this region of parameter space. Ahead of our comparisons to population level studies in \S \ref{sec:Discussion}, in Table~\ref{tab:temps+fluxes} we also quote X-ray luminosities, $L_{\rm X,0.1}$, in this band, as well as the ratio of those values to the bolometric luminosity of their respective star.

For 2MASS J0232-5746 (TIC 201861769), our \textit{XMM-Newton} results are compatible with those in Table~\ref{tab:temps+fluxes} from \textit{Swift}. The best fit model to the \textit{XMM-Newton} spectra had temperatures of $0.187^{+0.064}_{-0.049}$ and $0.745^{+0.133}_{-0.059}$\,keV, and emission measures of $\left(4.45^{+1.10}_{-0.44}\right)\times10^{50}$ and $\left(3.93^{+0.50}_{-0.88}\right)\times10^{50}$\,cm$^{-3}$, respectively. The temperatures agree excellently with \textit{Swift}, though for the lower temperature component, the emission measure is slightly lower in \textit{XMM-Newton}. The resulting best fit 0.3--2.4\,keV flux was also very slightly lower at $\left(5.86^{+0.18}_{-0.32}\right)\times10^{-14}$\,erg\,s$^{-1}$\,cm$^{-2}$, though within the measurement uncertainties from \textit{Swift}, demonstrating stability in the star's brightness across 10 years.

In Table~\ref{tab:elevatedFluxes}, we compare the fluxes of the defined elevated epochs to their respective star's quiescent level in the 0.3 -- 2.4\,keV band. Since these spectra are for the full length of elevated epochs, in the case of flares, these fluxes represent the average across the event rather than the peak. Despite that, in some cases, the average flux rose by an order of magnitude, far outshining the quiescent level. 

We also found archival, catalogued measurements of many of our low-mass targets in \textit{ROSAT} and \textit{eROSITA} data \citep{Boller2016,Merloni2024}. For ROSAT, we converted the catalogued count rates into fluxes using WebPIMMS\footnote{\url{https://heasarc.gsfc.nasa.gov/cgi-bin/Tools/w3pimms/w3pimms.pl}} with a single APEC temperature based on an average of the fitted XRT temperatures, weighted by their respective emission measures. For those from eROSITA, we multiplied the catalogued 0.2 -- 2.3\,keV flux by 0.85 to account for the one-size-fits-all power law model used to calculate the catalogued fluxes \citep[as per the method in][]{Foster2022}. We note, however, that this method assumes a temperature of 0.3\,keV, but since these stars are young, there is considerable emission around 1\,keV which may reduce the reliability of this method here. 

Despite this, the \textit{ROSAT} and \textit{eROSITA} fluxes were generally in very good agreement with those we measured with \textit{Swift}, with all but one within a factor of two of our XRT measurements. Small offsets such as this are unsurprising when factoring in slight differences in energy bands, analysis techniques, and the possibility of unidentified flares in the \textit{ROSAT} or \textit{eROSITA} data. The one star whose emission increased more significantly was UCAC4 204-001345, the star with the lowest flux and $L_{\rm X}/L_{\rm bol}$ in \textit{Swift}. Its 0.2-2\,keV flux in \textit{eROSITA} was $1.6\pm0.3$\,erg\,s$^{-1}$\,cm$^{-2}$, around four times the \textit{Swift} value, suggestive of a flare or other longer term variability of the emission. For the rest, the agreement between the telescopes is indicative of stable emission levels on timescales of years, up to decades in the case of those with \textit{ROSAT} detections.

\begin{table*}
\caption{The flare properties of the ten low-mass stars in our survey using data from Cycles 5, 6 (Sectors 56 -- 70) and Cycle 7 (Sectors 84 - 96). This includes the flare rates, the total flare number, the duration range and the energy range of the flares. We also include the rotation period of each star using data from Sectors 1 -- 70 (this work), the Rossby Number (this work), the stellar spectral types (from {\tt SIMBAD}) and the total number of stellar flares observed in all {\tess} Sectors from 1 -- 96. The stars have been split into their period bins using horizontal lines.}
   \begin{center}
   \label{tab:flare_properties}
\resizebox{1.0\textwidth}{!}{
	\begin{tabular}{lcccc|ccc|ccc|c}
	\hline
	           &          &     &     &             &  \multicolumn{3}{|c|}{Cycle 5} & \multicolumn{3}{|c|}{Cycle 7} \\
    \cline{5-10}

	TIC ID    & $P_{\rm rot}$ & $R_{o}$ & SpT & Flare Rate   & $\log(E)$         & Duration       & Flare Rate  & $\log(E)$       & Duration      & Total Flare    \\
	           &   (d)  & $\times10^{-3}$  &      & (per day)    & (erg)            & (min)          & (per day)   & (erg)          & (min)         & Number         \\
	\hline
    158596311 & 0.1547 & $1.32^{+0.44}_{-0.26}$ & 4.1 &  0.05 &  31.97           & 4.0            & 0.10        & 32.39 -- 33.46 & 8.0 -- 38.0   & 10    \\[0.1cm]
    425937691 & 0.1003 & $1.06^{+0.27}_{-0.18}$ & 5.5 &  0.24 &  32.00 -- 33.44  & 5.8 -- 47.9    & 0.26        & 32.14 -- 32.99 & 5.8 -- 24.0   & 60    \\
    \hline
    206544316 & 0.3220 & $4.2^{+1.4}_{-0.8}$ & 3.7 &  0.50 &  32.39 -- 34.25  & 5.8 -- 104.0   & 0.28        & 32.67 -- 33.42 & 5.8 -- 46.0   & 71    \\[0.1cm]
    156002545 & 0.3529 & $2.81^{+0.94}_{-0.56}$ & 3.4 &  0.15 &  32.07 -- 32.99  & 5.8 -- 20.0    & 0.29        & 32.09 -- 33.09 & 5.8 -- 87.9   & 14    \\
    \hline
    248354845 & 0.5219 & $3.55^{+0.89}_{-0.60}$ & 3.5 &  0.53 &  32.17 -- 33.82  & 13.3 -- 130.0  & --          & --             & --            & 47    \\[0.1cm]
    229142295 & 0.4472 & $1.80^{+0.60}_{-0.36}$ & 4.5 &  0.24 &  31.08 -- 31.47  & 4.0 -- 9.9     & 0.24        & 31.20 -- 32.62 & 5.8 -- 33.9   & 20    \\
    \hline
    220539110 & 0.7721 & $5.8^{+1.9}_{-1.2}$ & 3.0 &  0.37 &  32.27 -- 33.08  & 9.9 -- 20.0    & --          & --             & --            & 81    \\[0.1cm]
    166808151 & 0.7400 & $10.2^{+3.4}_{-2.0}$ & 2.0 &  0.92 &  32.14 -- 34.77  & 4.0 -- 120.0   & 1.08        & 32.11 -- 34.12 & 5.8 -- 40.0   & 135   \\
    \hline
    141807839 & 0.8471 & $6.1^{+1.5}_{-1.0}$ & 4.5 &  0.61 &  32.17 -- 34.45  & 4.0 -- 136.0   & 0.69        & 32.14 -- 34.94 & 5.8 -- 157.9  & 490   \\[0.1cm]
    201861769 & 0.8657 & $9.1^{+2.3}_{-1.5}$ & 4.1 &  0.05 &  33.85           & 38.0           & 0.14        & 31.98 -- 32.90 & 5.8 -- 24.0   & 10    \\
    \hline
    \end{tabular}}
    \end{center}
\end{table*}

\section{Optical Analysis \& Results}
\label{sec:tess_analysis}
Here, we discuss the independent analysis of the photometric optical light curves of {\tess}, including the refinement of the stellar rotation periods and stellar flare identification for each star.

\subsection{Refining Stellar Rotation Periods}

In Paper I, we used Cycle 1 and 3 {\tess} data to determine stellar rotation periods utilising a generalised Lomb–Scargle \citep[LS:][]{NRF1992, Zechmeister2009} periodogram for each of our targets. In this paper, we confirm these stellar rotation periods by combining all available {\tess} data from Cycles 1, 3, 5 and 7. We compute an LS periodogram, identifying the most prominent peak in frequency space to represent the rotation period of the star. We ensured the periods determined from a combination of all {\tess} data were consistent on each sector. The refined stellar rotation period values are provided in Table \ref{tab:flare_properties}. 

For TIC 425937691, the most dominant peak in the LS periodogram corresponds to a stellar rotational period of $P_{\rm{rot}}$ = 0.1~d, which is consistent with \citet{doyle2019, doyle2022}. However, in Sectors 95 and 96, there are additional significant peaks in the power spectra, including 0.067 d, and the rotation folded light curve using a period of 0.1 d shows a differing profile from those of sectors 68 and 69. This may indicate that additional spots have emerged by the time of the sector 95 and 96 observations.


\subsection{Stellar Flare Identification}
\label{flaretest}

Our method for identifying stellar flares is similar to that described in Paper I; however, we give a brief overview here for completeness.

We identify flares observed in {\tess} light curves for each star on a sector-by-sector basis, including Sectors 56 -- 96 (for previous Sectors, see Paper I). To do this, we implemented the open source {\sc Python} software {\sc AltaiPony} \citep{ilin2021flares} to detect and characterise flares in our sample automatically. In this analysis, we consider flares to consist of two or more consecutive points which lie above a mean threshold of greater than 2.5$\sigma$ \citep[see][]{davenport2014kepler, davenport2016kepler, doyle2018investigating, doyle2019}. Before any flare identification, each light curve was flattened, ensuring rotational modulation trends were removed. To do this, we utilised a Savitzky-Golay filter to detrend the light curve whose window length was defined so that it could resolve the periodic modulation but not remove short-duration events such as flares. For two of our targets, UCAC3~53-724 (TIC~425937691) and UPM~J0113-5939 (TIC~206544316), this was insufficient to remove the periodic signal; therefore, a custom detrending\footnote{\url{https://github.com/ekaterinailin/TESS_UCD_flares}} was utilised. This involved correcting global trends, fitting an LS periodogram to remove strong rotational modulation and masking and padding outliers. The light curve of each sector of data was manually vetted to confirm that only the rotational modulation was removed.

To determine how well we can recover flares in the {\tess} light curves, we used 
{\tt Altaipony} \citep{ilin2021flares} to inject and recover simulated stellar flares with typical profiles. We injected flares which had a duration between 4.3 min and 30 min and had a peak amplitude of at least 3 times the RMS of the flattened light curve after an outlier rejection to remove any flares. We were able to recover all injected flares over the whole range of the rotation period in our sample (0.10-0.87 d). The minimum flare energy which we could detect was largely set by the apparent brightness of the target. We are likely to miss flares with amplitudes lower than the above limit.

The resulting output from {\sc AltaiPony} includes many flare properties. For our analysis, we retrieve the start and stop times, flare amplitude and equivalent duration of each flare event. The start and stop times were used to determine the flare duration. To estimate energies for each flare in the {\tess} bandpass, we multiplied the equivalent duration by the quiescent luminosity (see Table \ref{tab:flare_properties} for all flare properties). Overall, we identified an additional 352 flares across the ten low-mass stars in our sample from Sectors 56 -- 96 with energies between 1.2$\times$10$^{31}$ and 8.7$\times$10$^{34}$~erg. This, combined with the flares from Paper I, gives a total of 938 flares from all available {\tess} data up to Cycle 7. In Figure \ref{fig:tess_lightcurves}, we plot the largest flare from AL 442 (TIC 141807839), which was observed in Sector 95 and reached a peak energy of 8.7$\times$10$^{34}$~erg. 

\subsection{Long-term Stellar Variability with {\tess}}

Since Paper I, our sample of low-mass UFRs has been observed in Cycle 5 (Sectors 56 -- 69) and Cycle 7 (Sectors 84 -- 96) with {\tess}. This provides a unique opportunity to compare their magnetic activity (i.e. through spot modulations and stellar flares) between Cycles 1, 3, 5 and 7, spanning from July 2018 to September 2025, with four years being directly observed. 

\begin{figure}
    \centering
    \includegraphics[width=0.97\linewidth]{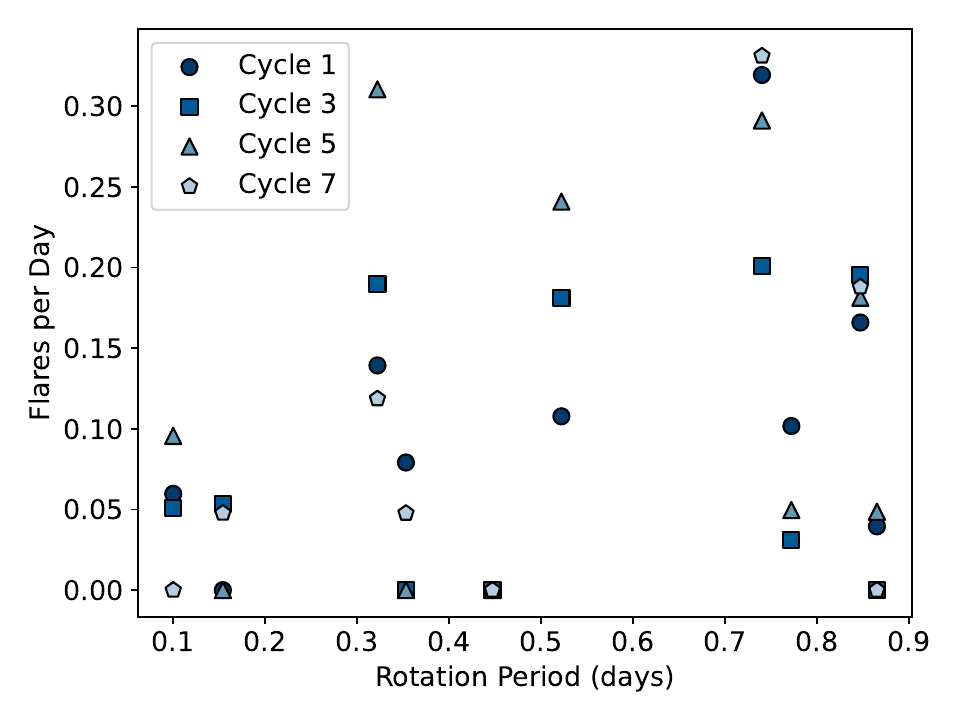} \\
    \includegraphics[width =0.97\linewidth]
    {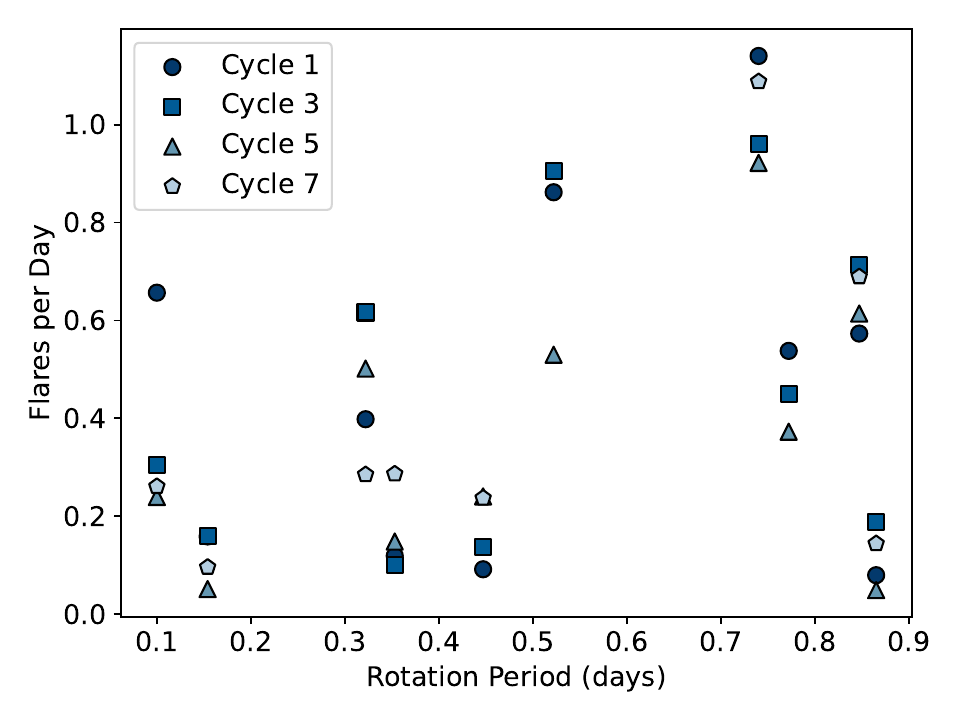}
    \caption{The flare number per day as a function of rotation period for each {\tess} Cycle 1 (circles), 3 (squares), 5 (triangles) and 7 (pentagons). {\it Top:} flares with energies > 10$^{33}$~erg only and {\it Bottom:} for all flares identified in each {\tess} Sector.
    \newline Note: Two targets did not have available SPOC data for Cycle 7. Additionally, there are potential targets where the flare rates in Cycles are comparable, leading to an overlap of plotted points.}
    \label{fig:flare_rates}
\end{figure}

In the lower panel of Figure \ref{fig:flare_rates}, we show the flare number per day as a function of rotation period for each of the stars in every observed {\tess} Cycle, which is an update of Figure 3 in Paper I. Our simulations using injected flares (\S \ref{flaretest}) indicated that given the cadence was the same for all observations we considered, the sensitivity to flare energy was similar over all sectors of data. However, in case there were some sectors with higher background, which might make the lowest energy flares more difficult to detect, we also show in the upper panel of Figure \ref{fig:flare_rates} the flare rate as a function of Cycle using only flares with energies >10$^{33}$~erg. We find evidence for a variable average flare rate using all flares and the more energetic flares.

Overall, for stars with P$_{\rm{rot}}$ > 0.6~days, we observe small changes in flaring activity rate on the order of 0.2 - 0.15 flares per day. However, for stars with $P_{\rm{rot}}$ < 0.6~d we notice an increased level of change in the flare rate per day, particularly amongst the three active stars in this period range. For TIC 425937691 (P$_{\rm{rot}}$ = 0.10~d), an initial high flare rate of 0.66 flares per day was observed in Cycle 1; however, over the next three cycles, this dropped significantly to ~0.3 flares per day, where it has remained since observations taken from August 2020. Additionally, TIC 206544316 (P$_{\rm{rot}}$ = 0.32~d) exhibits varying flare rates on the order of 0.3, which initially increased between Cycle 1 and 3, observed two years apart, before decreasing steadily over the next five years. In TIC 248354845 (P$_{\rm{rot}}$ = 0.52~d), a high flaring rate of 0.9 was observed in Cycles 1 and 3, which has dropped to 0.53 in Cycle 5, where observations were separated by 3 years. 

In addition to the flare rates per cycle, we searched through all available light curve data from Cycles 5 and 7 to identify changes in the morphology. With regards to the stellar rotation of our UFR low-mass stars, there is one standout candidate (TIC 206544316; UPM J0113-5939), which shows a change in the morphology of the light curve between sectors. Additionally, we also observe a factor of two change in the amplitude of the modulation. This is shown in Figure \ref{fig:tess_lightcurves}, for {\tess} Sectors 68 and 95, where in the latter sector a smoother rotational modulation shape is observed, along with an increased number of peaks. It is possible this could be a result of migrating spots to lower latitudes on the stellar surface, or the emergence of new active regions \citep[e.g.][]{Strassmeier2009, Roettenbacher2013, santos2019surface, ilin2021flares}. In \citet{bouma2023transient}, they identify TIC 206544316 as a Complex Periodic Variable (CPV) from their sample of 65,760 K and M dwarfs observed in 2-min cadence by {\tess} with T$_{\rm{mag}}$ < 16 and d <150~pc. They observed phases between ordinary sinusoidal modulation and highly structured modulation during Sectors 1 -- 55 of {\tess} data, which is similar to what we observed in later sectors. They conclude that CPVs are young ($\lesssim$150~Myr), low-mass ($\lesssim$0.4~\Msun) stars which display dips in flux with lifetimes of $\sim$100 cycles and attribute these to co-rotating clumps of gas or dust caught by the star's magnetic field \citep[see also][]{zhan2019complex}. 

\section{Discussion \& Conclusions}
\label{sec:Discussion}

In our previous studies \citep[see][]{doyle2019, ramsay2020ufr} we revealed a group of UFR M dwarf stars which displayed low levels of flaring activity in their {\tess} light curves. To explain this, in Paper I we investigated the presence of magnetic fields in ten UFR M dwarfs with P$_{\rm{rot}}$ < 1~day, utilising spectropolarimetric observations. Following on from this, in \citet{ramsay2022puzzling} we investigated the binarity of UFR low-mass stars using radial velocity measurements. Neither of these studies provided concrete explanations for the lack of flaring activity observed in UFR low-mass stars. Therefore, in this paper, we utilise {\sl Swift} and {\sl XMM-Newton} X-ray observations of the same M dwarfs from Paper I to determine their X-ray luminosities. We will now compare these to rotation-activity relations and determine what regime of emission our stars fall into. 

To test our sample for supersaturation observationally, we compare our $L_{\rm X,0.1}/L_{\rm bol}$ measurements from \S\ref{ssec:fluxes} to the sample of \citet{Wright2011,Wright2018}. In line with many other recent studies, they used Rossby number, the ratio of stellar rotation period and the convective turnover time \citep{Noyes1984}, on the x-axis as it provides a much tighter relation than using rotation period alone \citep[e.g.][]{Pizzolato2003,Wright2011}.

To estimate the convective turnover time of each star in our sample, we used the theoretical tracks from \citet{Landin2023}. For stars with a reliable age estimate, we used that to determine which position on the track to use. For the others, we used the sample on the track whose rotation period best matched the star. The tracks are also only gradated at 0.1\,$M_\odot$ intervals, so for those with a mass at or close to halfway between two tracks, we averaged across the two tracks on either side. \citet{Landin2023} estimated uncertainties of around five per cent on their values for Solar-like stars. As our sample is far outside of that regime, we adopted more conservative 20 per cent uncertainties on the convective turnover times, increasing to 25 per cent for stars lying between tracks and requiring averaging. In any case, the Rossby numbers calculated from these convective turnover times are all more than an order of magnitude below the value at which stars appear to transition from being saturated to unsaturated, and so we can be confident that all of our stars should either be in the former regime or show supersaturation.

\begin{figure}
    \centering
    \includegraphics[width=\columnwidth]{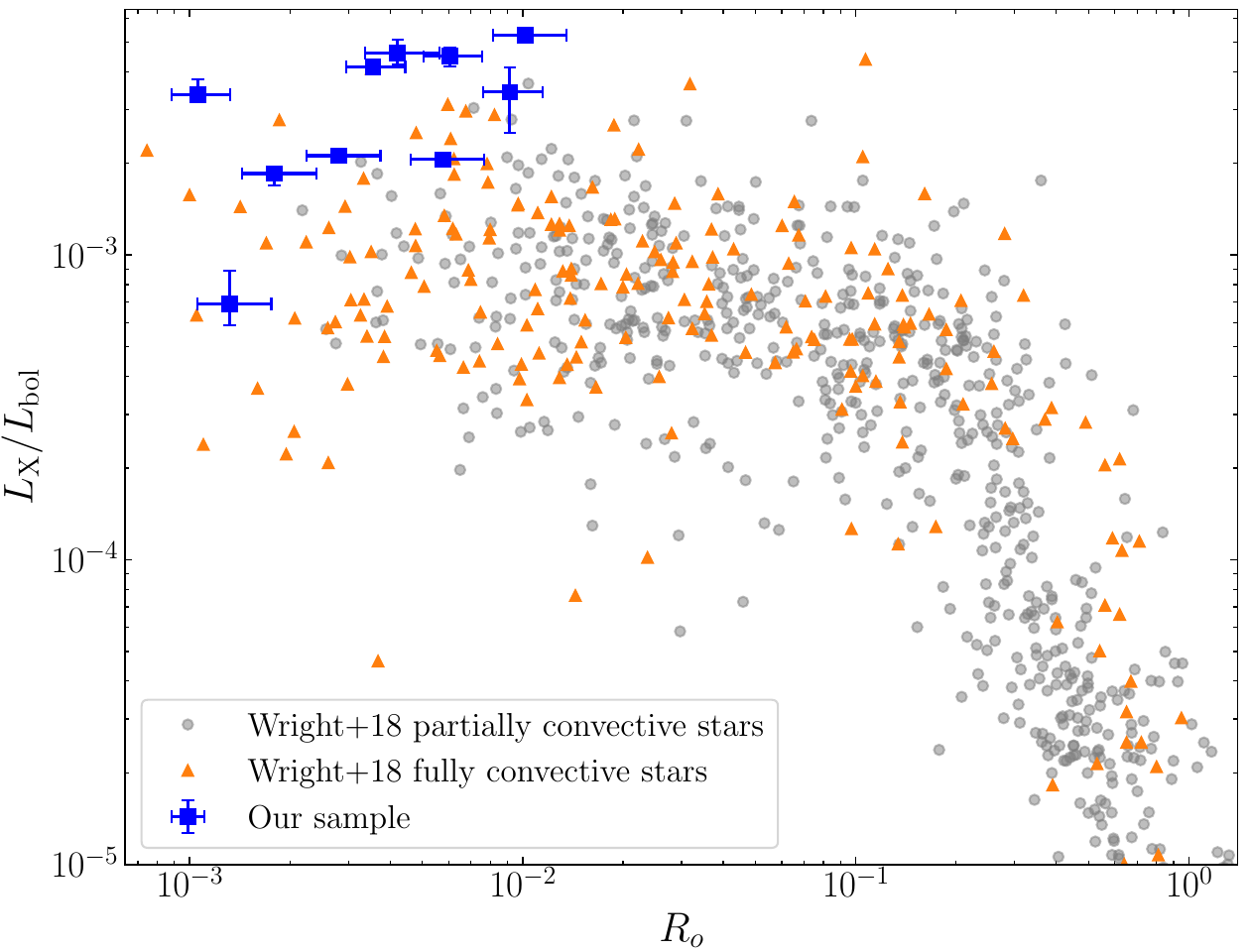}
    \caption{Comparison of our measured $L_{\rm X}/L_{\rm bol}$ values for the 0.1--2.4\,keV band to the sample of \citet{Wright2011,Wright2018}.}
    \label{fig:LxLbolRo}
\end{figure}

\begin{figure}
    \centering
    \includegraphics[width=\columnwidth]{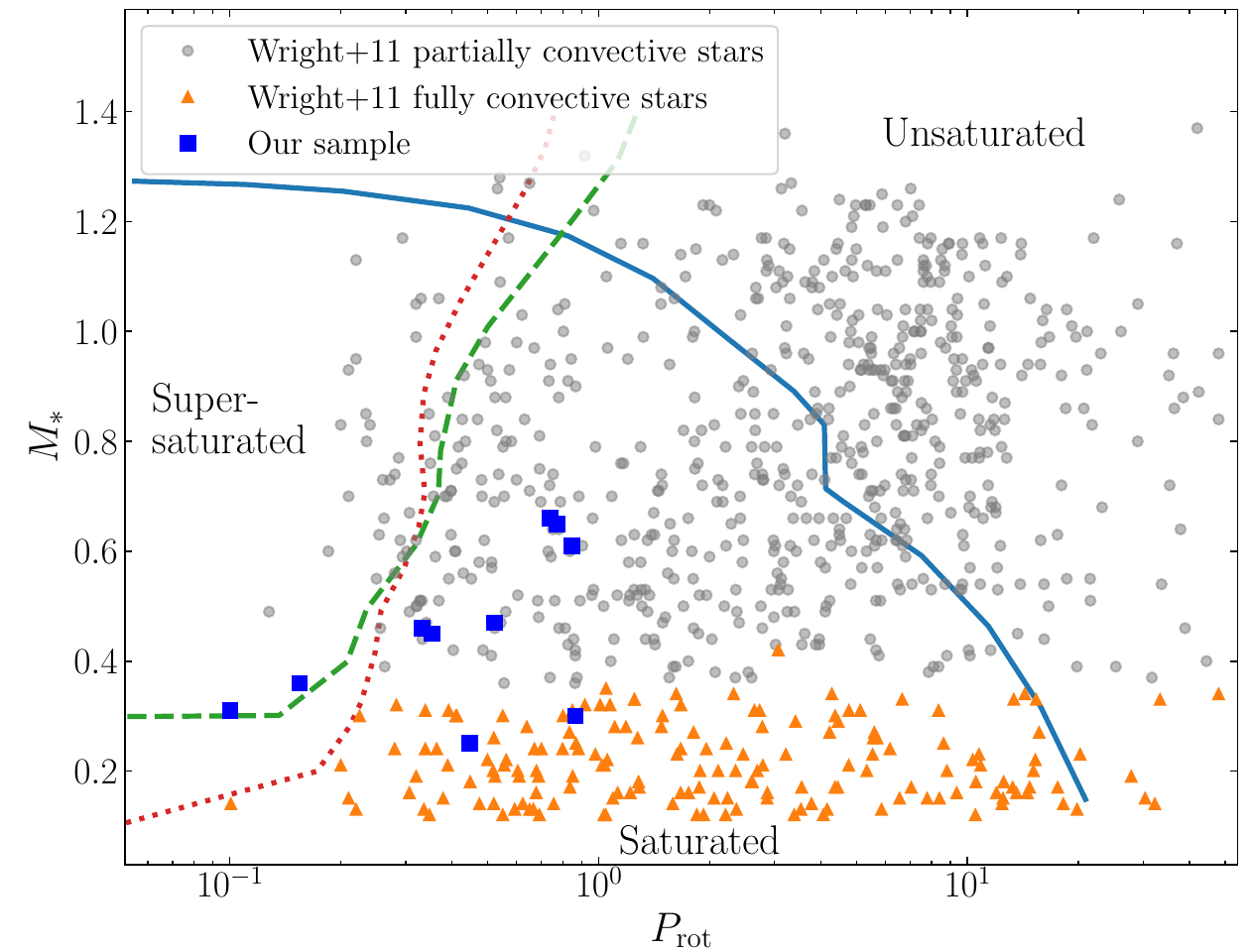}
    \caption{Stellar mass vs. rotation period for the stars in our sample and that of \citet{Wright2011,Wright2018}. We demonstrate the various regimes of emission with lines reproduced from Fig.~6 of \citet{Wright2011}: the solid blue line delineates the unsaturated regime from saturation, while two theoretical transitions to the supersaturated regime are shown with red dotted (centrifugal stripping) and green dashed (polar updraft) lines. We note that the positions of these supersaturation transitions may be different for pre-main sequence stars \citep[e.g][]{Argiroffi2016}.}
    \label{fig:massProt}
\end{figure}

In Figure~\ref{fig:LxLbolRo}, we plot $L_{\rm X,0.1}/L_{\rm bol}$ against Rossby number, $R_o$. This plot clearly demonstrates that all of the stars in our sample are at least at the expected level of a saturated star. UCAC4 204-001345 lies quite a bit below the other nine stars, but its emission is still in good agreement with the lower portion of the saturated regime. Its catalogued \textit{eROSITA} flux is also higher, though flaring cannot be ruled out as the cause. Therefore, as in \citet{jeffries2011investigating} and \citet{Argiroffi2016}, we see no evidence of supersaturation in the X-ray emission of our sample of UFR low-mass stars. Moreover, our results suggest that supersaturation is not the explanation for the relative flare inactivity observed in some of our sample.

To investigate this further, in Figure~\ref{fig:massProt} we show the stellar mass against rotation period for both our sample and those from \citet{Wright2011,Wright2018}, together with theoretical expectations for the location of the supersaturation region. The regime lines we have plotted are from Fig.~6 of \citet{Wright2011}, and are calculated for main-sequence stars. Only two of our stars, UCAC4 204-001345 and UCAC3 53-724, might be expected to be supersaturated based on this, but that is somewhat dependent on the stellar evolutional state and which mechanism generates supersaturation.. Based on their low mass and 45\,Myr age, one may expect both to still be pre-main sequence, and the polar updraft line (green dashed in Figure~\ref{fig:massProt}) is higher for pre-main sequence stars as calculated by \citet{Argiroffi2016}. These two stars would not be expected to be supersaturated in that instance. Therefore, there are no stars in our sample which are definitively expected to be supersaturated, which is compatible with our observational result of no supersaturation. 

In fact, rather than supersaturation, Figure~\ref{fig:LxLbolRo} appears to suggest that our sample may, on average, have somewhat higher values of $L_{\rm X}/L_{\rm bol}$ than the typical saturated star. Stellar system multiplicity, as discussed in Section~\ref{ssec:multiplicity}, is suggested for many of these systems and could potentially account for some of the apparent enhancement. The comparison \citet{Wright2011,Wright2018} sample will also have unresolved binaries too. Much of their sample was observed by \textit{ROSAT}, whose PSPC and HRI instruments would not have been able to resolve companions closer than a few arcsec apart at best. 
However, the \textit{Gaia} DR3 RUWE values for our sample appear to be systematically higher than those of the comparison sample. About two-thirds of the comparison sample have RUWE values less than 1.26, the lowest value of any of our sample stars. Multiplicity, therefore, likely plays a role in enhancing the apparent emission of our stars compared to expectations.

Other factors could play a role in enhancing $L_{\rm X}/L_{\rm bol}$ for some of our sample. Any companions enhancing the emission would be of roughly equal mass or lower, and as a result should only reduce the $L_{\rm X}/L_{\rm bol}$ of the individual components by a factor of about two (in the case of a binary; unresolved triples, quadruples etc, could be somewhat higher factors) versus the combined values in Table~\ref{tab:temps+fluxes} and Figure~\ref{fig:LxLbolRo}. This would still leave the vast majority of our sample at the very top of the saturated regime, far from supersaturation. Another possible contributing factor is that some of the youngest and/or lowest mass stars in our sample remain in the late pre-main sequence phase, for which slightly enhanced emission compared to young stars on the main sequence itself has been previously observed \citep{Argiroffi2016}.

\begin{figure}
    \centering
    \includegraphics[width=0.97\linewidth]{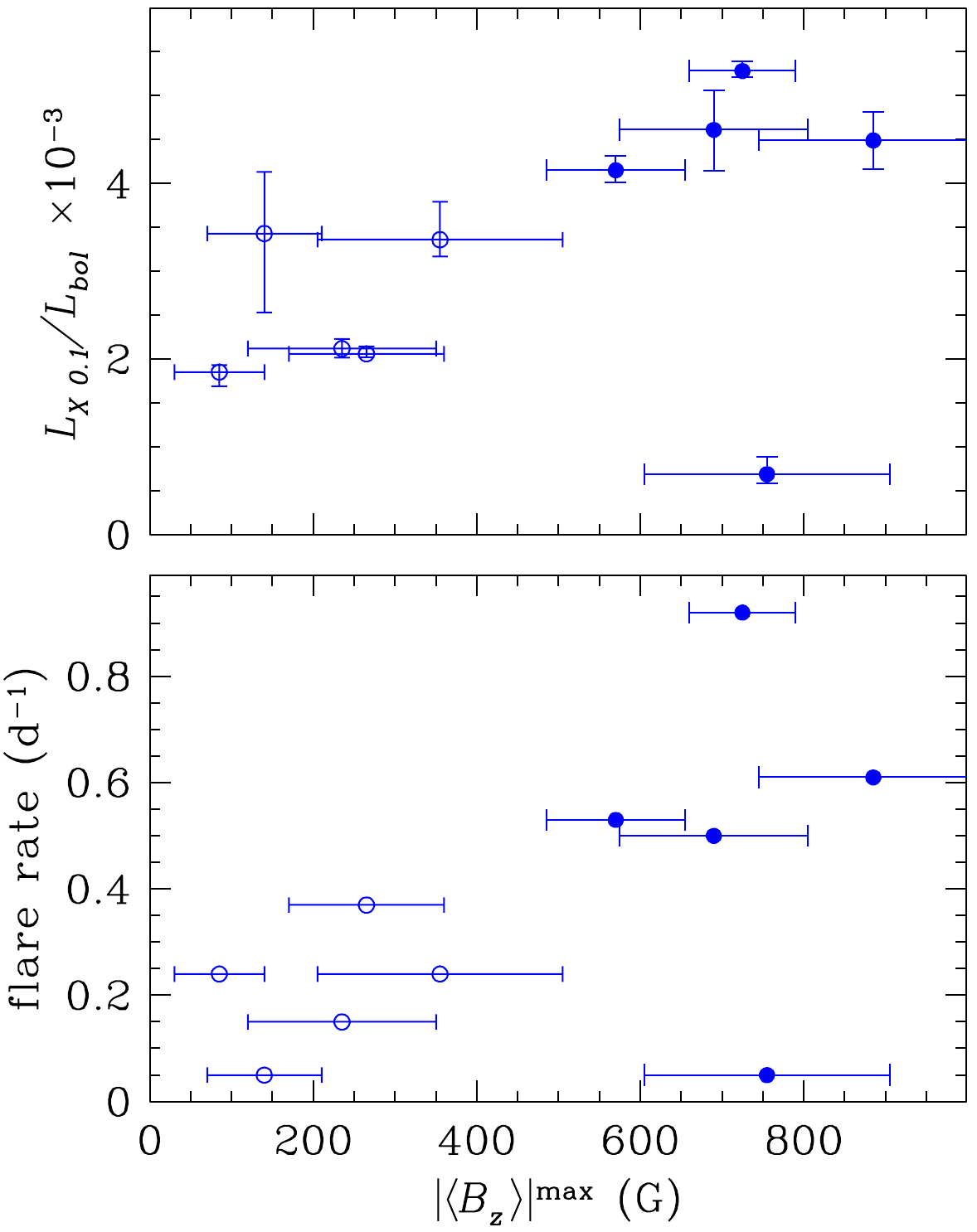}
    \caption{The magnetic field strength for each of our M dwarfs, derived from Paper I using spectropolarimetric observations from VLT/FORS2 (maximum GRISM 1028z values are used, taken from Table 3), plotted against the X-ray luminosity and flare rates. Solid circles represent stars where a field was detected, versus empty circles where a Bayesian approach determined the likelihood of the star having a detectable magnetic field was low (see \S4.1 of Paper I for full details).}
    \label{fig:Bz}
\end{figure}

In the top panel of Figure~\ref{fig:Bz}, we display our measured $L_{\rm X}/L_{\rm bol}$ values as a function of the magnetic field strengths we determined for the sample in Paper I. Intriguingly, four of the five stars in the sample whose magnetic fields were strong enough to be firmly detected are the four highest in terms of $L_{\rm X}/L_{\rm bol}$ in the sample. These same four are also all in the top five in terms of their detected XRT flux values. However, in addition to the aforementioned potential for unresolved multiplicity, the possible correlation between magnetic field strength and $L_{\rm X}/L_{\rm bol}$ for this sample 
is complicated by UCAC4 204-001345. It is the other star whose magnetic field was detected in Paper I, and yet has by far the lowest $L_{\rm X}/L_{\rm bol}$, and is also the dimmest in terms of raw XRT flux. A more definitive conclusion on this correlation would require much deeper constraints on system multiplicity, and perhaps on the magnetic field too, for those with no firm detection.

Considering the five period bins with pairs of optically high/low flaring stars, the more optically active star in each are all more X-ray active (both in terms of their $L_{\rm X}$ and $L_{\rm X}/L_{\rm bol}$) than their lower optically active counterparts. The five high optical activity stars are all in the top six for X-ray brightness overall. The low optical activity of 2MASS J0232-5746 barely pushes into fifth, though with uncertainties large enough that it could be much closer to the rest of the low optical activity stars. Altogether, this is highly suggestive of a correlation between optical and X-ray activity among this sample, though with the same multiplicity caveat we have mentioned previously.

In terms of X-ray flaring rates, we cannot conclude much quantitatively, as the \textit{Swift} data is far more sparse than {\tess} in terms of both cadence and integration time. The four stars with epochs designated as elevated, many of which are likely due to flaring, rank highly in terms of exposure time and/or have typically higher cadence data. It could simply be that we observed flaring events from those stars because they had the best chance of being detected due to how the observations were performed.

Our Sun is the only star for which the cycle of magnetic activity has been measured with great precision, resulting in a 22-year cycle, which consists of two 11-year sunspot cycles \citep{gnevyshev194822, hale1924sun}. To date, estimates of stellar activity cycles are known for dozens of stars from solar-type to low-mass cool dwarfs. These studies are based on spectroscopic observations of Ca II H \& K lines as well as photometric light curves \citep[e.g.][]{baliunas1995chromospheric, vida2014looking, reinhold2017evidence, ramsay2024searching}. Early-M dwarfs (M0 - M3) exhibit magnetic activity cycles on timescales of a few to tens of years, although these cycles are typically less regular and shorter than the solar cycle \citep[e.g.][]{da2011long, mascareno2016magnetic}. In contrast, fully convective mid to late M dwarfs (M4 onwards) show little evidence for coherent long-term activity cycles, with their variability instead dominated by stochastic flaring and rapid starspot evolution \citep[e.g.][]{robertson2013halpha, davenport2015detecting}. 

With regards to our sample of ten UFR low-mass stars, we investigated the changing flare rate per day as a proxy for potential stellar activity cycles. We find a relatively low change in flare rate for stars with P$_{\rm{rot}}$ > 0.6~d, which agrees with previous findings of a decrease in flare activity for low-mass stars as their rotation period decreases \citep[also linked to age][]{davenport2019evolution}. In Figure \ref{fig:Bz}, we also show the relationship between flare rate from Cycle 5 and the magnetic field strength determined in Paper I. We observe a higher field strength for the more flare active stars which also corresponds to the stars which had a higher likelihood for a detectable magnetic field (as with the X-rays in the top panel, UCAC4 204-001345 is an exception). For three of our stars with P$_{\rm{rot}}$ < 0.6~d, we find evidence of long-timescale modulation within flare rates over an $\sim$7 year period during {\tess} Cycle 1, 3, 5 and 7 observations. This could indicate potential magnetic activity cycles within these stars; further observations will be required to observe a change in the flare rates of these stars. However, given the expected cycle length, continued observations in the X-ray and optical bands are required over many years. Currently observing Sector 101, {\tess} will continue to take observations well into 2028 and continue to provide long baseline observations of stars, making stellar activity cycle studies possible. 

\section*{Acknowledgements}
LD would like to acknowledge funding from the UK Space Agency. We include data in this paper collected by the {\tess} mission, where funding for the {\tess} mission is provided by the NASA Explorer Program. This work presents results from the European Space Agency (ESA) space mission Gaia. Gaia data are being processed by the Gaia Data Processing and Analysis Consortium (DPAC). Funding for the DPAC is provided by national institutions, in particular the institutions participating in the Gaia MultiLateral Agreement (MLA). The Gaia mission website is \url{https://www.cosmos.esa.int/gaia}. The Gaia archive website is \url{https://archives.esac.esa.int/gaia}. This work made use of data supplied by the UK Swift Science Data Centre at the University of Leicester. We also used observations obtained with \textit{XMM-Newton}, an ESA science mission with instruments and contributions directly funded by ESA Member States and NASA.

\section*{Data Availability}
All {\tess} 2-minute and SPOC light curves are available from the NASA MAST portal. Gaia DR2 and DR3 data are available from the Gaia Archive. {\sl Swift} data are available through either the NASA HEASARC archive or the UK \textit{Swift} Science Data Centre. The {\sl XMM-Newton} data is available from the \textit{XMM-Newton} Science Archive.



\bibliographystyle{mnras}
\bibliography{UFR_xray} 

\appendix

\section{X-ray spectral plots}

\begin{figure*}
    \centering
    \includegraphics[width = 0.97\linewidth]{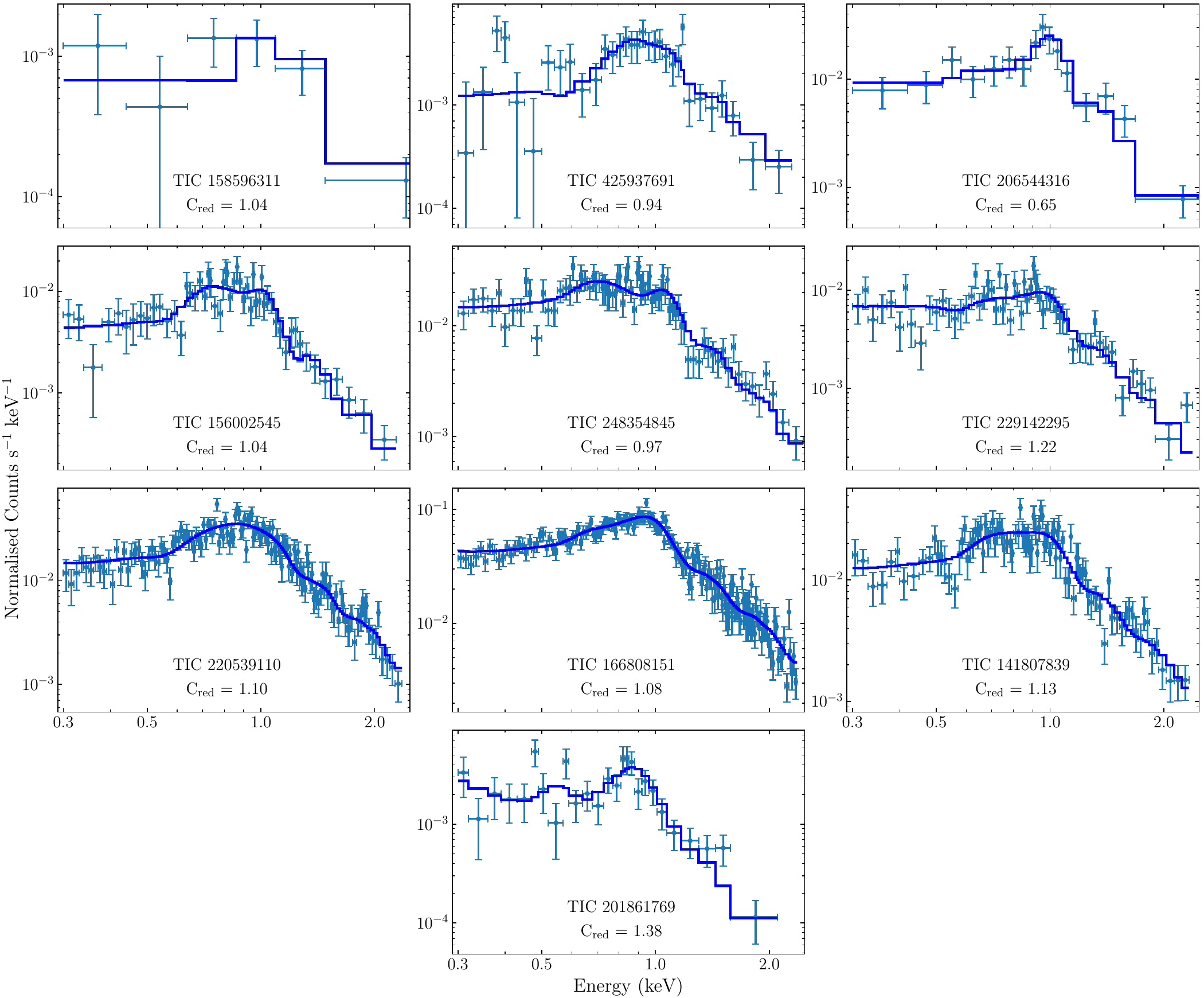}
    \caption{\textit{Swift} XRT spectra for each of our stars. Overplotted is the best fit model (see \S \ref{ssec:xSpectra}). For each star we give the reduced C-statistic, C$_{\rm red} =$ C-statistic/degrees of freedom.}
    \label{fig:swiftSpec}
\end{figure*}

\begin{figure*}
    \centering
    \includegraphics[width = 0.48\linewidth]{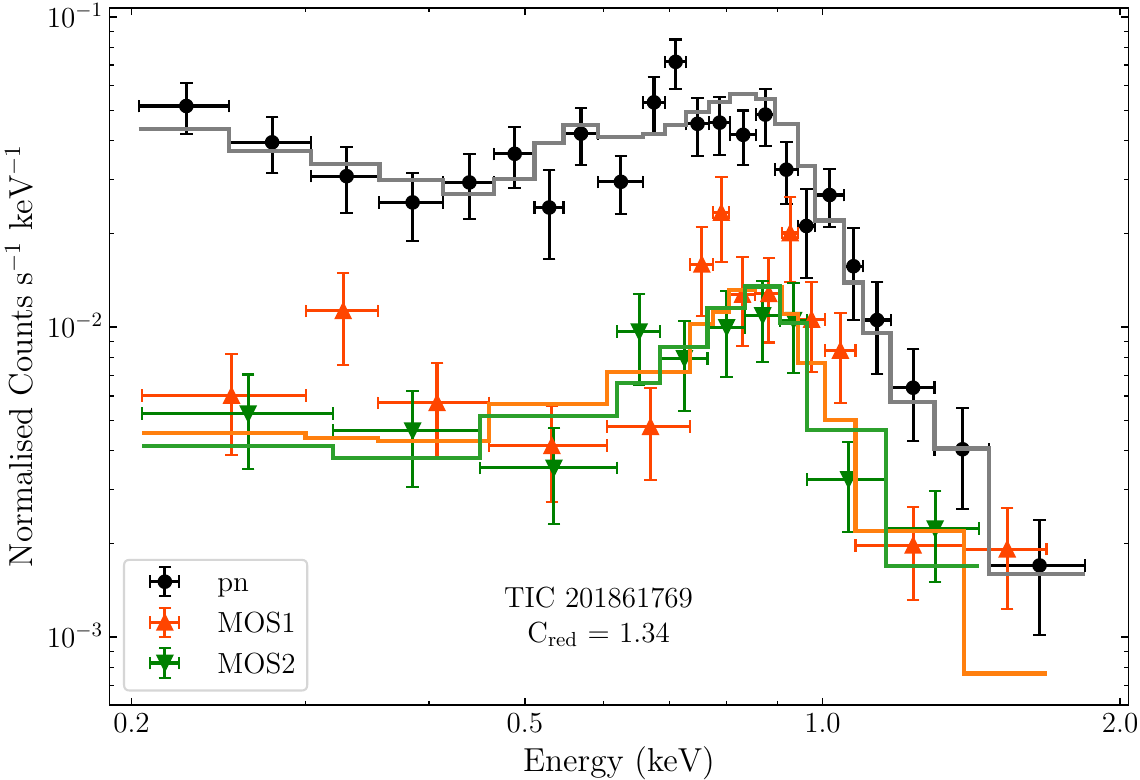}
    \caption{\textit{XMM-Newton} spectra for 2MASS\,J0232-5746 from each of the three EPIC detectors. Overplotted is the best fit model (see \S \ref{ssec:xSpectra}).}
    \label{fig:XMMspec}
\end{figure*}

\section{{\tess} light curves of UPM J0113-5939 and AL 442}

\begin{figure*}
    \centering
    \includegraphics[width = 0.97\linewidth]{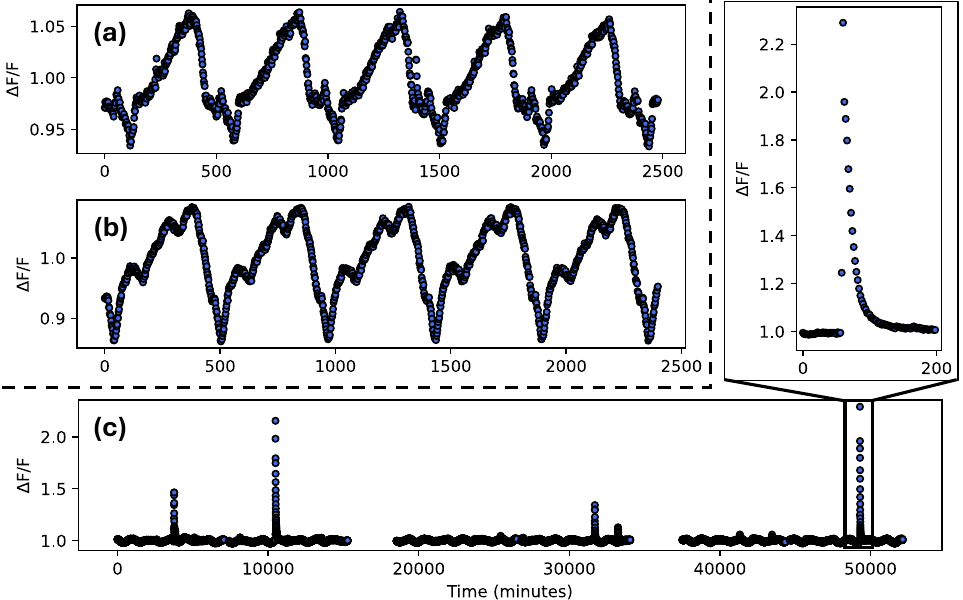}
    \caption{A selection of {\tess} light curves from two of our low-mass UFRs. The top two plots show the evolving rotational modulation shape for UPM J0113-5939 (TIC 206544316) from Sector 68 (a) and Sector 95 (b). The bottom plot (c) shows the flaring activity for AL 442 (TIC 141807839) in Sectors 94 \& 95, with the inset image zoomed on the largest flare in our sample at 8.7$\times$10$^{34}$~erg. The dashed line has been added to separate the plots and make it clear that these are from two different stars.}
    \label{fig:tess_lightcurves}
\end{figure*}

\bsp	
\label{lastpage}
\end{document}